\documentclass[twocolumn]{aastex63}

\received{--}
\revised{--}
\accepted{--}

\submitjournal{ApJ}

\shorttitle{Alfv\'{e}nic waves in inclined coronal loops}
\shortauthors{Skirvin et al.}

\usepackage{amsmath}
\usepackage{mathtools}

\graphicspath{{./}{figures/}}

\begin{document}

\title{Alfv\'{e}nic motions arising from asymmetric acoustic wave drivers in solar magnetic structures}

\correspondingauthor{Samuel Skirvin}
\email{samuel.skirvin@kuleuven.be}

\author[0000-0002-3814-4232]{Samuel J. Skirvin}
\affiliation{Centre for mathematical Plasma Astrophysics, Department of Mathematics, KU Leuven, Celestijnenlaan 200B bus 2400, B-3001 Leuven, Belgium}

\author[0000-0002-6641-8034]{Yuhang Gao}
\affiliation{Centre for mathematical Plasma Astrophysics, Department of Mathematics, KU Leuven, Celestijnenlaan 200B bus 2400, B-3001 Leuven, Belgium}
\affiliation{School of Earth and Space Sciences, Peking University, Beijing, 100871, People’s Republic of China}

\author[0000-0001-9628-4113]{Tom Van Doorsselaere}
\affiliation{Centre for mathematical Plasma Astrophysics, Department of Mathematics, KU Leuven, Celestijnenlaan 200B bus 2400, B-3001 Leuven, Belgium}

\begin{abstract}
Alfv\'{e}nic motions are ubiquitous in the solar atmosphere and their observed properties are closely linked to those of photospheric p-modes. However, it is still unclear how a predominantly acoustic wave driver can produce these transverse oscillations in the magnetically dominated solar corona. In this study we conduct a 3D ideal MHD numerical simulation to model a straight, expanding coronal loop in a gravitationally stratified solar atmosphere which includes a transition region and chromosphere. We implement a driver locally at one foot-point corresponding to an acoustic-gravity wave which is inclined by $\theta = 15^{\circ}$ with respect to the vertical axis of the magnetic structure and is similar to a vertical driver incident on an inclined loop. We show that transverse motions are produced in the magnetic loop, which displace the axis of the waveguide due to the breaking of azimuthal symmetry, and study the resulting modes in the theoretical framework of a magnetic cylinder model. By conducting an azimuthal Fourier analysis of the perturbed velocity signals, the contribution from different cylindrical modes is obtained. Furthermore, the perturbed vorticity is computed to demonstrate how the transverse motions manifest themselves throughout the whole non-uniform space. Finally we present some physical properties of the Alfv\'{e}nic perturbations and present transverse motions with velocity amplitudes in the range of $0.2-0.75$ km s$^{-1}$ which exhibit two distinct oscillation regimes corresponding to $42$ s and $364$ s, where the latter value is close to the period of the p-mode driver in the simulation.
\end{abstract}

\keywords{Magnetohydrodynamics (1964); Solar atmosphere (1477); Solar chromosphere (1479); Solar oscillations (1515); Solar coronal loops (1485); Solar coronal waves (1995); Magnetohydrodynamical simulations (1966)}

\section{Introduction} \label{sec:intro}
Observations of the solar corona over the last few decades have revealed that transverse oscillations are ubiquitous throughout \citep{nak1999, Tomczyk_et_al_2007, McIntosh_et_al_2011, Morton2015, Morton2019}. These motions are commonly interpreted as magnetohydrodynamic kink waves, due to the displacement of the axis of the observed magnetic waveguide \citep{erd2007,van_etal08b}. However, some authors have favoured the term `Alfv\'{e}nic' to classify the observed oscillations as coupled MHD wave modes in an inhomogeneous environment \citep[e.g.][]{goo2009}. It is widely believed that Alfv\'{e}nic waves, and their associated energy dissipation mechanisms, may contribute to the problems of coronal heating and acceleration of the solar wind \citep[e.g. see review by][]{vanDoorsselaere2020SSRv}. Excellent recent reviews on Alfv\'{e}nic waves in a solar context can be found in \citet{Banerjee2021} and \citet{Morton2022}.

Solar p-modes are globally coherent standing resonant acoustic waves, generated by turbulent convection within the convection zone, with periods that possess peak power of around $5$ minutes. These waves are driven by broadband acoustic noise as a result of photospheric granulation, hence they possess wave periods corresponding to the rough lifetime of a granule. Any typical location on the solar surface is oscillating with an amplitude of a few hundred m s$^{-1}$ \citep{Ulrich1970} which results in about $10$ cm s$^{-1}$ per individual mode \citep{Priest2014}. Solar p-modes have been shown to leak power into the lower atmosphere, which may be important to explain some observed dynamics through driving jet phenomena \citep{dep2004, Hansteen2006, Heggland2007, Skirvin2023_JET}  and may be used as a tool for seismology of the local plasma \citep{Chaplin2008}. 

Typical oscillation periods, in both open and closed magnetic field configurations, measured in the corona possess peaks at three to five minutes \citep{DeMoortel2002,VanDoorsselaere2008_EISCORONA, Tomczyk2009, Morton2016, Uritsky2021, Gao2022}, which suggests a connection to photospheric p-modes with the same peak period. Recently, \citet{Morton2019} have provided direct evidence of this, using data from the Coronal Multi-channel Polarimeter (CoMP) coronagraph, and suggest that the additional flux provided by p-mode conversion in the solar atmosphere should be incorporated in wave heating models, where instead it is commonly ignored. However, the waves excited by p-modes are predominantly acoustic in nature and face many challenges in their propagation to the upper atmospheric layers, including the acoustic cut-off and wave reflection resulting from gravitational stratification. Whilst our understanding of acoustic wave propagation has been developed through theories such as the `ramp'-effect \citep{Bel1977, Jeffries2006}, mode conversion \citep{Schunker2006, Cally2017, Riedl2019} and mode absorption \citep{Cally2000, Cally2003, Hindman2008}, the relationship between coronal Alfv\'{e}nic waves and their corresponding p-mode frequency spectra is still unclear.

In reality, the lower solar atmosphere comprises of a partially ionised plasma which introduces non-ideal MHD effects on the conversion of acoustic to magnetic MHD waves. This occurs either through direct interactions of charged particles with the magnetic field or via collisional coupling between charged particles and neutrals. While our aim in this study is to investigate wave conversion and generation in ideal MHD, it is worth briefly mentioning the non-ideal effects which are expected to be present in the lower solar atmosphere. Firstly, the Hall effect, which arises due to a drift between electrons and ions, has been shown to introduce a new coupling mechanism through the presence of the Hall current \citep{Cally2015, Gonzalez-Morales2019}, which may be responsible for coupling fast magnetoacoustic and Alfv\'{e}n waves in the case where there is no cross-field wave propagation. Additionally, ambipolar diffusion can effectively dissipate Alfv\'{e}n wave energy \citep{Khomenko2019,Ballester2020} when wave frequencies are sufficiently high (on the order of $>0.1$ Hz) \citep{cally2019}. Recently, \citet{soler2021} have demonstrated that these non-ideal effects are able to sustain the chromospheric heating rate in the foot-points of coronal loops. 

In this study,  our primary aim is to investigate whether p-modes inclined to the vertical axis of a magnetic structure can produce the observed Alfv\'{e}nic perturbations in the corona, within a stratified and structured solar atmosphere. In Section \ref{sec:methods} we introduce the numerical model for the simulations. The results on the excited modes in the simulation are presented in Section \ref{sec:results}, while the observational signatures of such waves are provided in Section \ref{sec:obs_features}. Finally, we present a discussion and conclusions of our results in Section \ref{sec:conclusions}.

\section{Methods} \label{sec:methods}
\subsection{Model} \label{subsec:model}

Following the initial setup for the simulations of \citet{Riedl2021}, we use the same coronal loop model of \citet{Reale2016} featuring a straightened, evacuated loop spanning from photosphere to photosphere in a cylindrical coordinate system. In other studies it is common practice to model a coronal loop as a density enhancement dictated by some smoothly varying density profile. However, in our case, the loop is `evacuated' in the sense that it is not defined by a density enhancement, instead, the radial structuring is provided through an increase in the magnetic field strength. The original thermodynamic model is adapted from \citet{Serio1981} and the initial equilibrium is obtained by numerically relaxing a hydro-static model over time, featuring a photosphere, transition region and corona, with vertical and straight magnetic field, as done in \citet{Guar2014}. However, as a result of increased plasma and magnetic pressure at the centre of the loop, the magnetic field ultimately expands in the corona during the equilibration process. At the foot-points, the magnetic field reaches a maximum of $273$ G which decreases to $13$ G at the loop apex at $z = 0$ Mm, which is appropriate for quiet Sun conditions of the network field. Furthermore, the total magnetic field decreases with radial distance across the loop and its magnitude has a profile which is approximately Gaussian-shaped. To avoid repeating information on this process in this work, the full description of how the equilibrium is obtained can be found in \citet{Riedl2021}. 

Gravity is incorporated into the model in the form:
\begin{equation}
    g(z)=-g_\odot \cos{\left(\frac{(z-z_0)\pi}{L}\right)},
\end{equation}
where $g_\odot=274$ m s$^{-2}$ is the gravitational acceleration at the solar surface, $z_0$ is the $z$-coordinate of the photosphere located at the base of our numerical domain, and $L = 61.61$ Mm is the total length of the loop. Taking a gravitational profile in this form results in a gravitational acceleration acting in the negative $z$-direction for $z<0$ and to the positive $z$-direction for $z>0$ with $g(0)=0$ m s$^{-2}$ at the loop apex. Gravity across the loop, as well as the loop curvature, are neglected.

\begin{figure}
    \centering
    \includegraphics[width=0.48\textwidth]{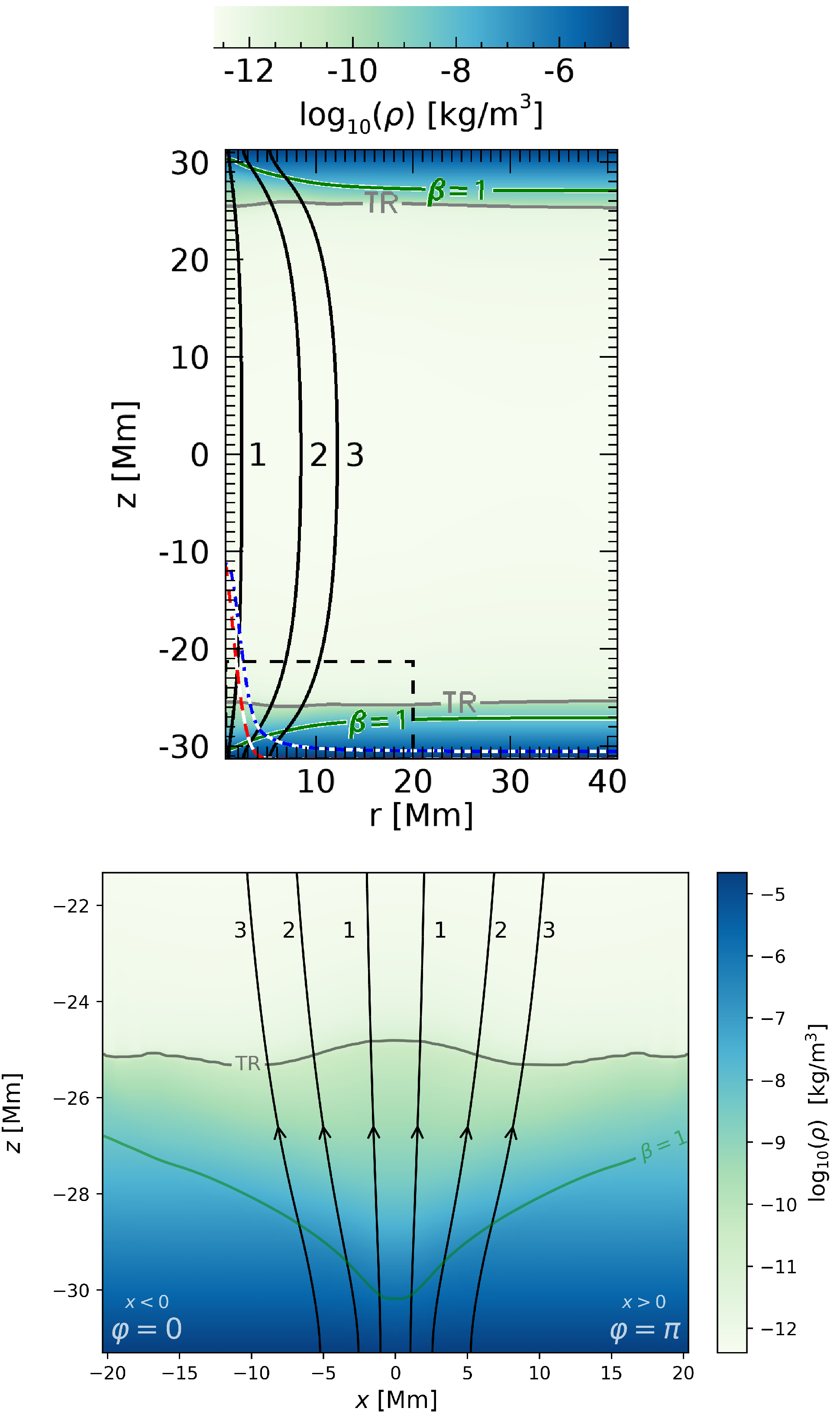}
    \caption{Full simulation domain taken from \citet{Riedl2021} (top) showing the equilibrium density. Also shown is the location and strength of the driver (red dashed line) and the strength of the magnetic field (blue dashed line), both in arbitrary units. A zoomed-in region delineated by the black dashed box is displayed in the bottom panel. This figure shows two azimuthal cuts taken at $\varphi=0$ corresponding to $x<0$ and another cut at $\varphi=\pi$ denoted by the region $x>0$. The colour shading depicts the background plasma density and selected labelled magnetic field lines are shown by the solid black lines. The location of the transition region (grey curve) and the $\beta=1$ layer (green curve) are also marked.}
    \label{fig:atmosphere}
\end{figure}

The initial model is presented in Figure \ref{fig:atmosphere} which shows the plasma-$\beta = 1$ contour along with the position of the transition region (TR) which, for simplicity, is defined as the location where the plasma temperature $T = 40,000$ K, however as the TR has a finite width, the TR contour should be regarded as an approximation. We also indicate the labels of specific field lines, which are chosen such that field line $1$ is rooted inside the magnetic loop at the foot-point, whereas field line $2$ is rooted at the full-width-half-maximum (FWHM) of the total magnetic field strength. Therefore, field line $3$ is rooted in a weaker region of magnetic field and also positioned outside the local wave driver location (see Section \ref{subsec:driver}). These field lines will be used in subsequent analyses to study how the radial structuring of the plasma affects wave propagation.

\subsection{Numerical Setup and Boundary Conditions}\label{subsec:setup}

Taking advantage of the azimuthal symmetry of our setup to reduce computing time, only one half of the loop cylinder is simulated. This differs from the simulations of \citet{Riedl2021}, where only one quarter of the loop due to the symmetry of both the equilibrium and of the driver. Our simulation domain ranges from $0.73$ Mm to $41.01$ Mm in the radial direction, from $0$ to $\pi$ in the $\varphi$-direction, and from $-31.31$ Mm to $31.31$ Mm in the $z$-direction, with $192$ × $256$ × $768$ data points, respectively. The loop axis is actually located at $r=0$ however we do not simulate close to this region due to the regular singularity at the origin of the domain. The numerical mesh is stretched, in the radial and vertical directions, in certain regimes of the domain to ensure that increased resolution is obtained closer to the enhanced region of magnetic field, whereas decreased resolution is utilized in regions of low stratification such as the upper corona and at large radial distances away from the center of the magnetic loop. An exact description of the numerical grid and resolution in different regimes can be found in \citet{Riedl2021}. The mesh and resulting resolution are uniform in the azimuthal direction.

The simulations are performed using the PLUTO code \citep{Mign2007, Mign2012, Mign2018}, where the ideal MHD equations are solved in 3D cylindrical coordinates using the Harten-Lax-Van Leer (HLL) approximate Riemann solver, with a piece-wise total variation diminishing (TVD) linear reconstruction method for the spatial integration. We utilize reflective boundary conditions for both boundaries in the $r$-direction and anti-symmetric boundary conditions in the $\varphi$-direction, where the signs for the tangential components of the magnetic field and velocity field are reversed. Additionally, we incorporate anti-symmetric boundaries for the upper $z$ boundary. At the lower $z$ boundary, the same boundary conditions are set; however the velocity, pressure and density are perturbed according to an analytical solution for a gravity-acoustic wave, given by the description in Section \ref{subsec:driver}.

\subsection{Vector component decomposition}
As previously mentioned, a cylindrical coordinate system is used for the simulation employed in this work. However, due to the expansion of the magnetic field in the chromosphere and corona, we do not have a purely vertical magnetic field and it would be useful to decompose vector components respective to the background magnetic field vector.

In full 3D simulations, where the magnetic field is not confined to a 2D geometry, the isolation of MHD waves becomes non-trivial as there are an infinite number of vectors normal to the magnetic field vector \citep{Yadav2022}. To help distinguish between the different types of waves in our simulation, we adopt a similar decomposition method to that used in \citet{Riedl2019}. The conversion of components from a cylindrical geometry ($r, \varphi, z$) to those parallel, perpendicular and azimuthal to magnetic flux surfaces is given by:
\begin{eqnarray}
    \mathbf{e}_{\parallel} &=& \left[\frac{B_r \text{cos}(\varphi)}{\sqrt{B_r^2 + B_z^2}}, \frac{B_r \text{sin}(\varphi)}{\sqrt{B_r^2 + B_z^2}}, \frac{B_z(\varphi)} {\sqrt{B_r^2 + B_z^2}}\right], \label{decomp_parallel} \\
    \mathbf{e}_{\varphi} &=& [-\text{sin}(\varphi), \text{cos}(\varphi), 0], \label{decomp_azi} \\
    \bf{e}_{\perp} &=& \bf{e}_{\varphi} \times \bf{e}_{\parallel}, \label{decomp_perp}
\end{eqnarray}
where $\mathbf{e}$ denotes a unit vector in each direction, respectively. Equations (\ref{decomp_parallel})-(\ref{decomp_perp}) set up a Cartesian basis describing the vector decomposition with respect to magnetic field lines for an equilibrium magnetic field which is structured in the $r$ and $z$ directions in a cylindrical geometry. The component of magnetic field azimuthal to magnetic surfaces is ignored in the decomposition due to the field lines being circularly symmetric around the axis of the loop and no magnetic twist is considered in the initial model. This decomposition of components parallel, perpendicular and azimuthal to the magnetic field lines will be important in the context of understanding the wave modes which are present in the simulation.

\subsection{Driver}\label{subsec:driver}

The gravity-acoustic wave driver implemented in this work perturbs all components of the velocity vector as well as the plasma pressure and density taking the form \citep{Mihalis1984, Khomenko2012, Santamaria2015}:
\begin{eqnarray}
    v_r' = &A& |v| \exp \left(\frac{z}{2H}+\Im(k_z)z\right)\cos(\varphi) \nonumber \\ &\times&\sin \left( \omega t - \Re(k_z)z -k_{\perp}r \cos(\varphi)+\phi_v \right), \label{eq:driver_velocityr}\\    
    v_{\varphi}'= &-A& |v| \exp \left(\frac{z}{2H}+\Im(k_z)z\right)\sin(\varphi) \nonumber \\ &\times&\sin \left( \omega t - \Re(k_z)z -k_{\perp}r \cos(\varphi)+\phi_v \right), \label{eq:driver_velocityphi}\\            
    v_z'= &A& \exp \left(\frac{z}{2H}+\Im(k_z)z\right) \nonumber \\ &\times&\sin \left( \omega t - \Re(k_z)z -k_{\perp}r \cos(\varphi)\right), \label{eq:driver_velocityz}\\
    p' = &A& |P| \exp \left(\frac{z}{2H}+\Im(k_z)z\right) \nonumber \\ &\times&\sin \left( \omega t - \Re(k_z)z -k_{\perp}r \cos(\varphi) + \phi_P\right), \label{eq:driver_pressure}\\
    \rho' = &A& |\rho| \exp \left(\frac{z}{2H}+\Im(k_z)z\right) \nonumber \\ &\times&\sin \left( \omega t - \Re(k_z)z -k_{\perp}r \cos(\varphi) + \phi_{\rho}\right), \label{eq:driver_density}
\end{eqnarray}
where,
\begin{eqnarray}
    |v| &=& \frac{c_s^2 k_{\perp}}{\left( \omega^2 - c_s^2 k_{\perp}^2\right)} \sqrt{\Re(k_z)^2 + \left(\Im(k_z) + \frac{\gamma - 2}{2H\gamma} \right)^2}, \label{eq:v_amp}\\  
    |P| &=& \frac{\gamma \omega}{\left( \omega^2 - c_s^2 k_{\perp}^2\right)} \sqrt{\Re(k_z)^2 + \left(\Im(k_z) + \frac{\gamma - 2}{2H\gamma} \right)^2}, \label{eq:P_amp}\\  
    |\rho| &=& \frac{\omega}{\left( \omega^2 - c_s^2 k_{\perp}^2\right)} \nonumber \\ &\times& \sqrt{\Re(k_z)^2 + \left(\Im(k_z) - \frac{1}{2H} + \frac{\gamma - 1}{H\gamma}\frac{k_{\perp}^2c_s^2}{\omega^2} \right)^2}, \label{eq:rho_amp}\\       
     k_r &=& \cos(\varphi) \sin(\theta)\left|\frac{\omega}{c_s}\sqrt{\frac{(\omega_c^2 -\omega^2)}{(\omega_g^2 \sin(\theta)^2-\omega^2)}}\right|, \label{eq:kr}\\   
     k_{\varphi} &=& \sin(\varphi) \sin(\theta)\left|\frac{\omega}{c_s}\sqrt{\frac{(\omega_c^2 -\omega^2)}{(\omega_g^2 \sin(\theta)^2-\omega^2)}}\right|, \label{eq:kphi}\\   
     k_{\perp} &=& \sqrt{k_r^2 + k_{\varphi}^2}, \label{eq:kperp}\\ 
     k_z &=& \sqrt{\frac{\omega^2 - \omega_c^2}{c_s^2} - k_{\perp}^2\frac{\omega^2-\omega_g^2}{\omega^2}}, \label{eq:kz}  \\
     \omega_c &=& \frac{c_s}{2H}\cos(\theta_B),
     \label{eq:omega_c} \\   
     \omega_g &=& \frac{2\omega_c\sqrt{\gamma-1}}{\gamma}, \label{eq:omega_g} \\
     \phi_P &=& \arctan\left(\frac{\Im(k_z) + (\gamma-2)/2 H\gamma}{\Re(k_z)}\right), \label{eq:phi_p} \\
     \phi_v &=& \phi_P, \label{eq:phi_v} \\
     \phi_{\rho} &=& \arctan\left(\frac{\Im(k_z) -\frac{1}{2H} + \frac{(\gamma-1)(c_s^2 k_{\perp}^2/\omega^2)}{\gamma H}}{\Re(k_z)}\right). \label{eq:phi_rho}
\end{eqnarray}
Here, $A=300$ m s$^{-1}$ is the driver amplitude, which agrees with photospheric Doppler oscillations from the contribution of p-modes \citep{McClure2019}. The relative amplitudes for the velocity, pressure and density perturbations are denoted as $|v|$, $|P|$ and $|\rho|$, respectively; $H$ is the pressure scale height; $k_z$ is the vertical wavenumber, which only has a real part in our case; $\omega = 2\pi/T$ is the driver frequency, with period $T = 370$ s, which is within the typical range of p-mode periods. The variables $\phi_v$, $\phi_P$ and $\phi_{\rho}$ are the velocity, pressure and density phase shifts compared to the vertical velocity perturbation, $\hat{v}_z$. The acoustic cut-off and thermally modified acoustic cut-off frequencies are denoted as $\omega_c$ and $\omega_g$, respectively, where $\theta_B$ is the angle between the magnetic field and direction of gravitational acceleration. It is worth noting that the perpendicular wavenumber $k_{\perp}$ of the driven waves in this simulation is constant for all azimuthal angles and all times. This is because of the sinusoidal dependence on the azimuthal angle $\varphi$ for $k_{\varphi}$ and the cosine dependence on the azimuthal angle for $k_r$, resulting in a constant upon adding the square of these values. The angle of the driver with respect to the vertical axis is represented by $\theta$ and is taken to be $\theta = 15^{\circ}$ in this work, note that this is similar to a vertical driver incident on a coronal loop which itself is inclined to the vertical axis. Since we consider the field-aligned component of gravity, an inclined coronal loop would result in an appropriate reduction of the gravity near the loop foot-points. For the inclination angle considered in this study, this results in roughly a $3\%$ correction to the field-aligned gravity, which is a very limited modification to the inclined driver modeled herein. It is evident from Equations (\ref{eq:kr}) and (\ref{eq:kphi}) that when the inclination of the driver, or the loop axis, is removed ($\theta=0$), then only waves with a vertical wavenumber are excited, this specific case study was discussed in detail by \citet{Riedl2021}.

The driver is applied locally at only one foot-point at the base of the domain, and located inside the loop. This is achieved by multiplying Equations (\ref{eq:driver_velocityr})-(\ref{eq:driver_density}) by the function:
\begin{equation}
    D(r) = \exp\left( - \frac{r^2}{\sigma^2}\right),
\end{equation}\label{eq:Gaussian_driver}
where $\sigma$ is the standard deviation of the Gaussian describing the width of the driver, which in this study is taken to be $\sigma = 2$ Mm. At the photospheric base of the domain, the FWHM of the Gaussian magnetic field strength is located at $r = 2.58$ Mm, therefore the driver can be considered to be applied within the foot-point loop radius.

It is important to note that the initial equilibrium contains background velocities of up to $18$ km s$^{-1}$ \citep{Reale2016, Riedl2021}, therefore, two simulations are run, one without the implementation of the wave driver and one simulation with the driver. As a result, the perturbed quantities can be recovered by subtracting the simulation without a driver from the simulation with the effects of the driver.

\section{Results} \label{sec:results}
\subsection{Wave modes excited}\label{subsec:motions}
We wish to investigate if an inclined gravity-acoustic wave driver can excite higher order modes of a magnetic cylinder. \citet{Riedl2021} have shown that a purely vertical driver excites tube waves which are axisymmetric, when the background is also axisymmetric, corresponding to the $m=0$ sausage mode of a magnetic cylinder. This result is not entirely unexpected and may explain the axisymmetric perturbations of structures observed in the lower solar atmosphere \citep{mor2012, moreels2015, Gao2021,grant2022}. However, with an inclined driver, we expect that tube waves which are non-axisymmetric in nature, i.e. kink waves, should also be excited within the magnetic structure. Determining whether these modes are present in the simulation is important as kink/Alfv\'{e}nic motions with peak power associated with p-modes are readily seen in observations \citep{Tomczyk_et_al_2007, Morton2019, Gao2022} and they may carry significant energy to the corona. 

\subsubsection{Non-axisymmetric motions}\label{subsec:axis_motion}

One of the main distinguishing properties between the $m=0$ and $\lvert m\rvert=1$ modes of a magnetic cylinder is the perturbation of the axis of the waveguide. The $\lvert m\rvert=1$ mode is the only mode (in a uniform magnetic flux tube model) which perturbs the central axis of the magnetic flux tube \citep{edrob1983}, whereas the $m=0$ and $\lvert m\rvert >2$ modes do not perturb the axis of the structure. As our simulation domain does not start at $r=0$ (due to the regular singularity at this location), we instead choose to convert the vectors to a Cartesian geometry and look at the $\hat{v}_x$ component. We adopt a `hat' notation to refer to perturbed quantities representing those taken from the simulation with the driver minus the simulation without a driver. After converting to the Cartesian grid, we interpolate the $\hat{v}_x$ signal and take a slice which corresponds to the position of the axis at $x=0$.
\begin{figure*}
\gridline{\fig{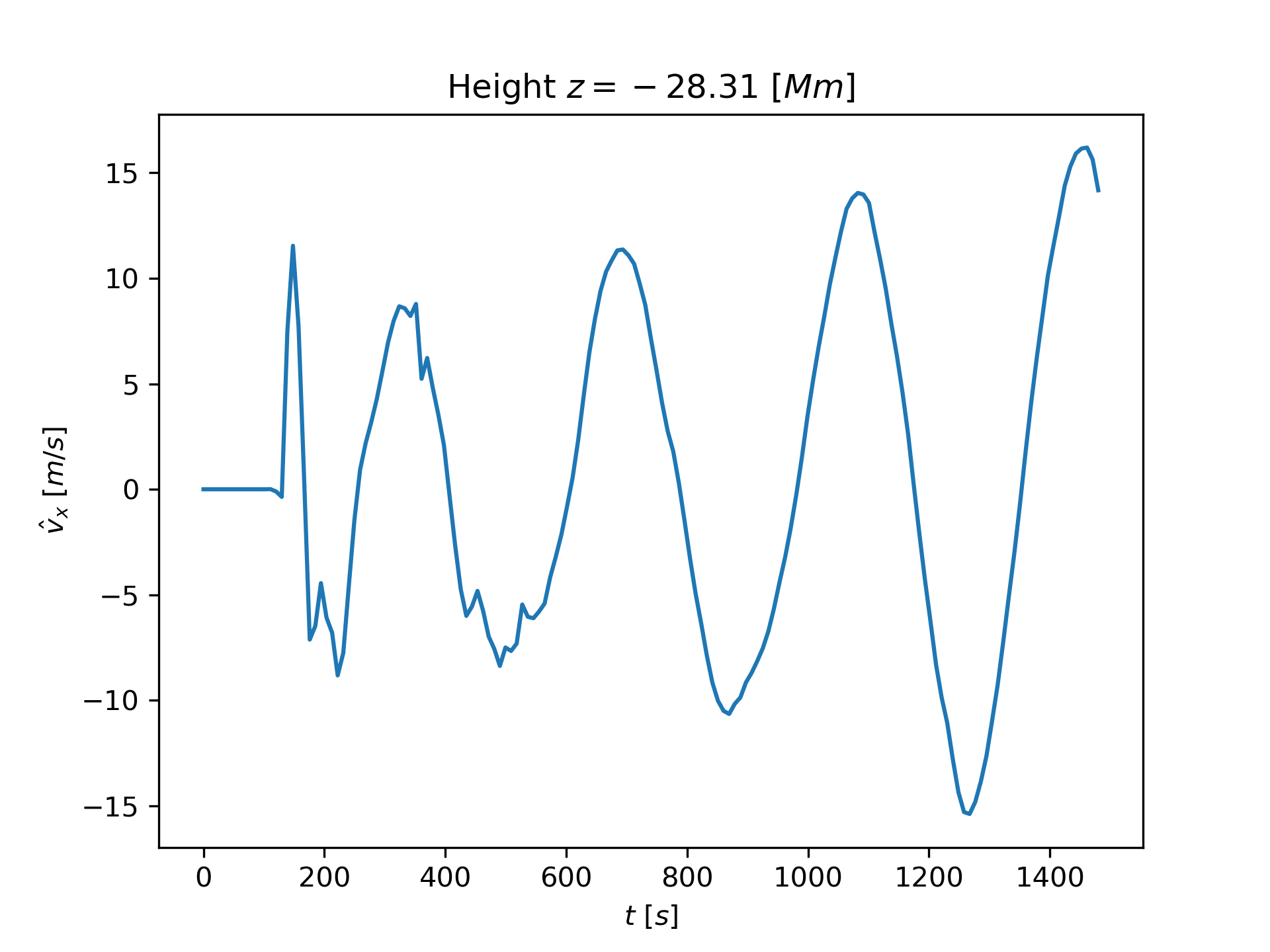}{0.45\textwidth}{(a)}
          \fig{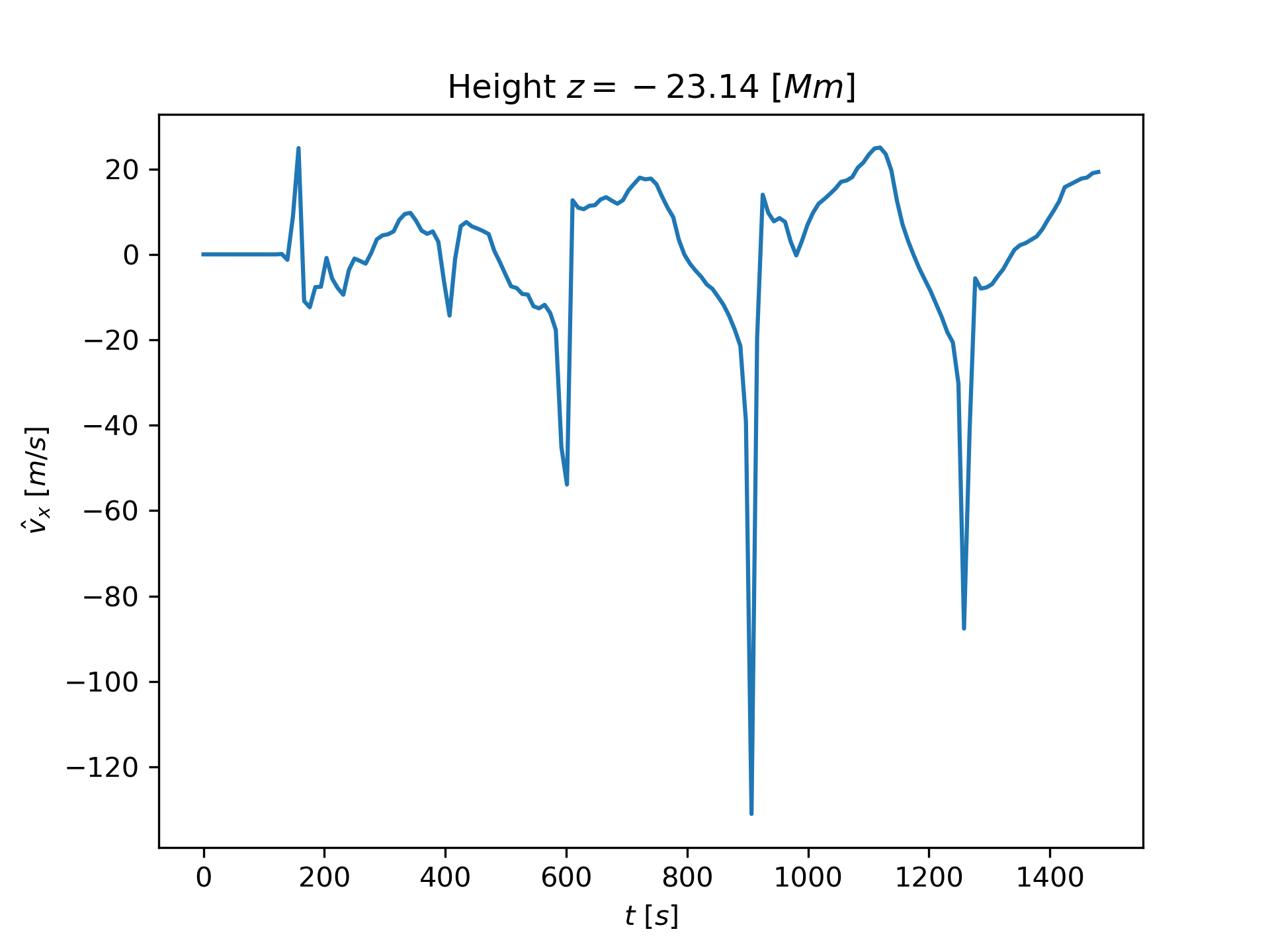}{0.45\textwidth}{(b)}
          }          
\caption{The component of $\hat{v}_x$, after interpolation, taken at a slice corresponding to the centre of the loop axis ($[x,y]=[0,0]$) as a function of time over the simulation in (a) the photosphere and (b) above the transition region in the corona.
\label{fig:axis_motions}}
\end{figure*}
Shown in Figure \ref{fig:axis_motions} are the resulting $\hat{v}_x$ amplitudes at two different heights in the simulations corresponding to photospheric and coronal heights, respectively. For both heights, there is a clear periodic oscillation of the $\hat{v}_x$ amplitude at the loop axis, denoting the movement of the axis of the structure, indicating the presence of non-axisymmetric waves. The photospheric signal displays a smooth sinusoidal signal with a period corresponding to that of the driver, with the amplitude of the axis displacement growing in time. On the other hand, whilst the coronal signal still displays a periodic behaviour, the signal is more distorted, which may be a result of non-linear interactions with the transition region. The asymmetry present in the amplitude of the coronal signal between positive and negative values is a result of the asymmetry of the inclined driver in a preferred azimuthal angle, resulting in a $\hat{v}_x$ signal which is also asymmetric, this behaviour is more pronounced in the corona which can be attributed to the density stratification. Nonetheless, there is clear evidence that an inclined acoustic-gravity wave driver, mimicking that of a p-mode, can indeed displace the axis of a magnetic structure in the solar atmosphere, however we should investigate in more detail the exact dominance of the wave modes excited.

Additionally, non-axisymmetric motions of a magnetic cylinder can also be determined by analysing the radial and azimuthal velocity perturbations. For example, modes of a magnetic cylinder with $\lvert m\rvert \geq 1$ produce azimuthal velocity perturbations ($\hat{v}_{\varphi}$), whereas the $m=0$ mode does not produce any azimuthal perturbations to either the velocity or magnetic field.
\begin{figure*}
\gridline{\fig{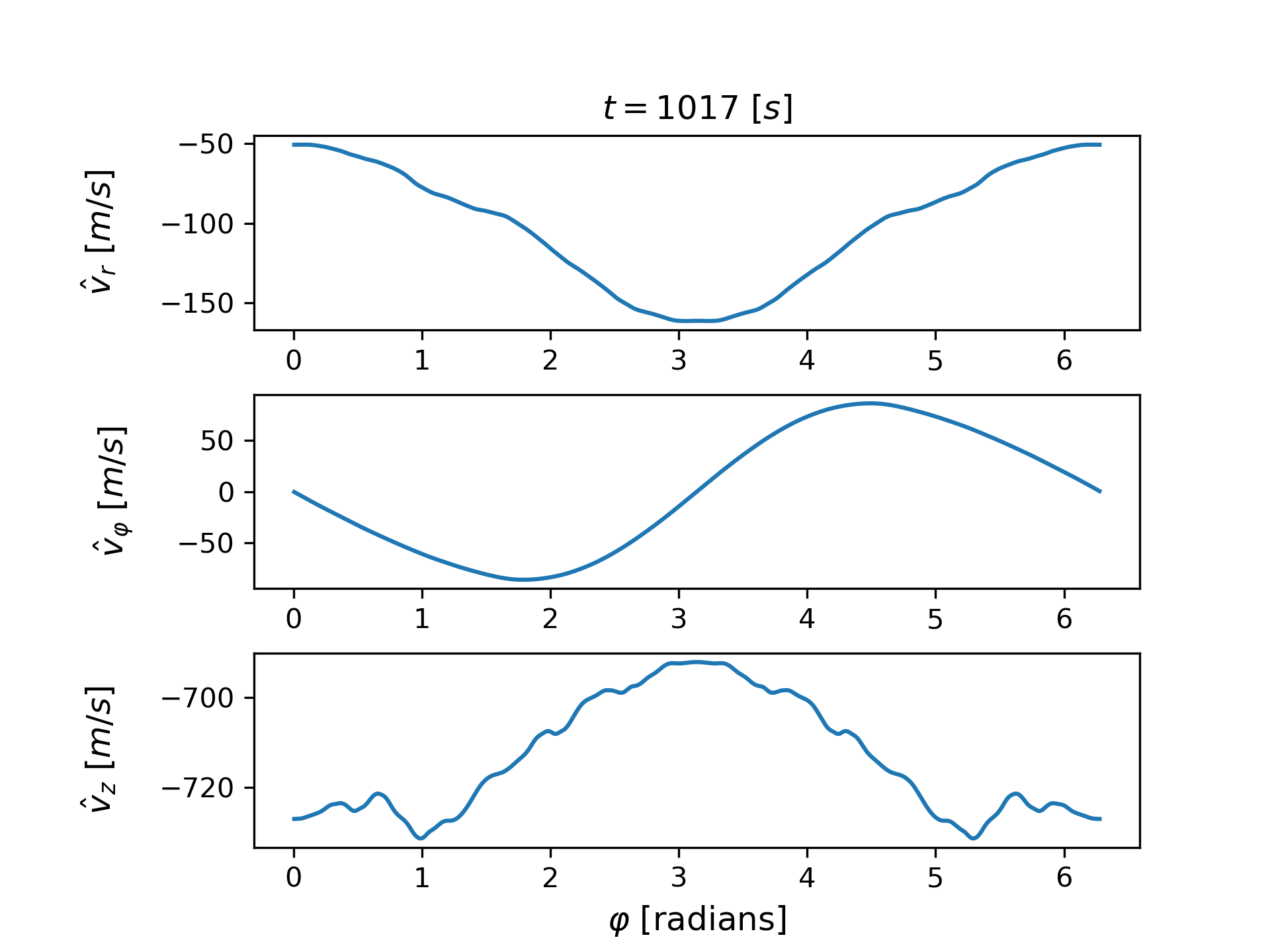}{0.49\textwidth}{(a) Photospheric signals}
          \fig{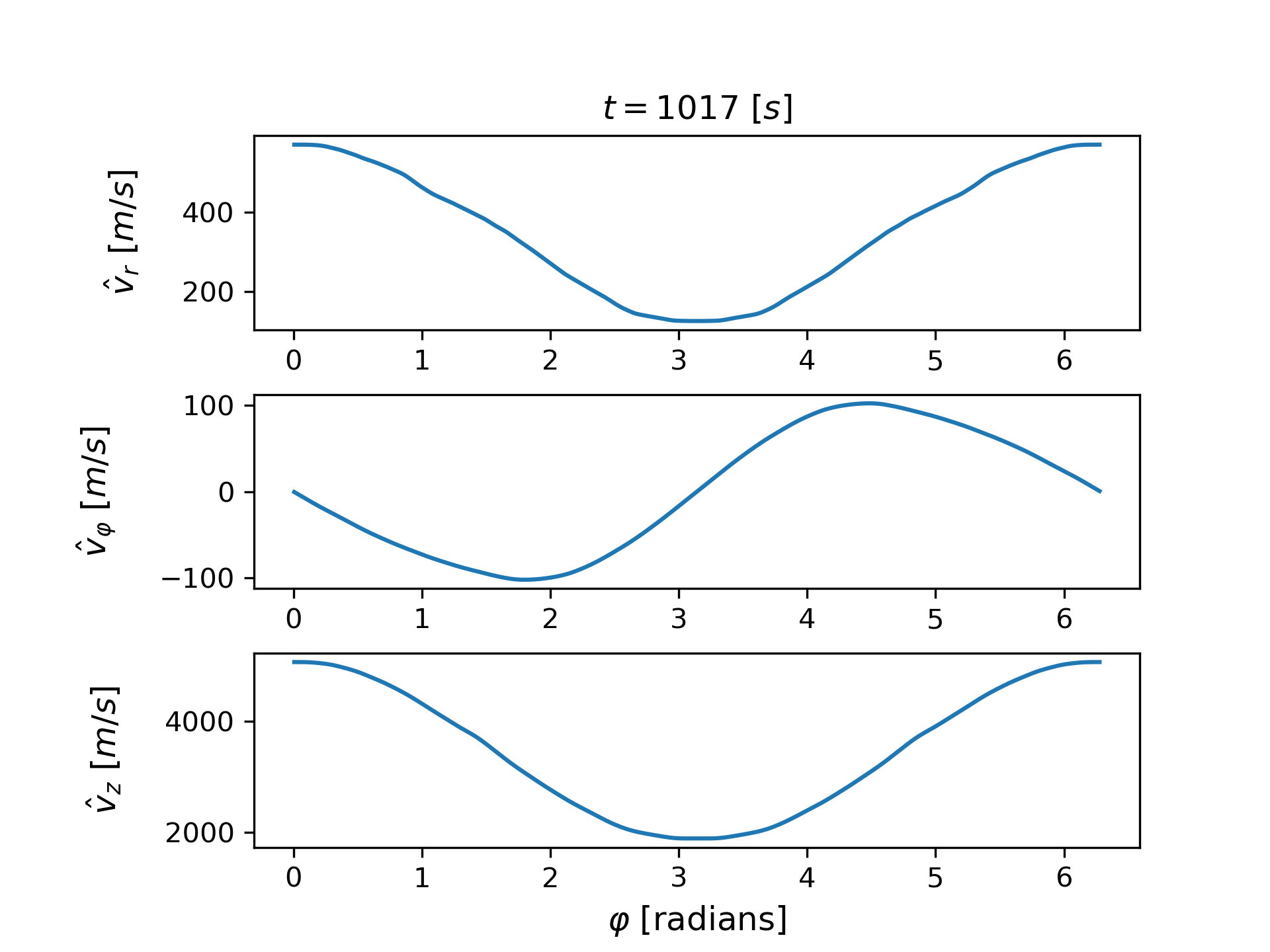}{0.49\textwidth}{(b) Coronal signals}
          }          
\caption{The signals of $\hat{v}_r$, $\hat{v}_{\varphi}$ and $\hat{v}_z$ over the flux surface corresponding to a point on field line $1$ in (a) the photosphere and (b) the corona as a function of azimuthal angle $\varphi$ for a given point in time ($t=1017$ s) in the simulation. The signals have been extended to cover a full period ranging from $0$ to $2\pi$ in the azimuthal direction.
\label{fig:vr_vphi_signals}}
\end{figure*}
Figure \ref{fig:vr_vphi_signals} shows the resulting signals for the radial, azimuthal and vertical perturbations of the velocity at a specific time snapshot in the simulation. The signals are produced at a radial distance which corresponds to a point lying on field line $1$ in both the photosphere and the corona (see Figure \ref{fig:atmosphere}). The signals in both the photosphere and the corona display similar characteristics, with the only noticeable difference being the amplitude of the perturbations is larger in the corona, which is expected as a result of the vertical density stratification. The similarity of the signals in the azimuthal direction can be attributed to the uniformity of the background model in this direction. The existence of a $\hat{v}_{\varphi}$ component suggests the presence of higher order ($\lvert m\rvert \geq 1$) modes present in the simulation. We can see that the $\hat{v}_{\varphi}$ signals display a sinusoidal behaviour with respect to the azimuthal angle of the magnetic loop, whereas the $\hat{v}_{r}$ signals are cosine in appearance. This is the expected behaviour of the classical kink motion of a cylindrical flux tube undergoing a transverse motion, which is a result of the asymmetric inclined p-mode driver. On the other hand, the $\hat{v}_r$ signal also provides information on the $m=0$ sausage motions present in the magnetic structure. If we were to expect a pure sausage mode perturbation, then the signal of $\hat{v}_r$ would be constant as a function of $\varphi$, oscillating in amplitude over time, whereas the $\hat{v}_{\varphi}$ signal would be zero. It can be clearly seen that the $\hat{v}_r$ signal in the azimuthal direction is not constant, hinting at the existence of higher order modes in the loop.

It should be noted here that the amplitudes of the radial and azimuthal velocity perturbations are still significantly smaller than the vertical component, $\hat{v}_z$, within which the perturbation from the driver dominates. Also evident in Figure \ref{fig:vr_vphi_signals} is the effect of the inclined driver on the $\hat{v}_z$ component in favour of the azimuthal angle $\varphi=0$. For an angle of $\varphi=0$, this component possesses an absolute maximum, which decreases as the azimuthal angle approaches $\varphi=\pi$.

\subsubsection{Fourier analysis of azimuthal wavenumbers}\label{subsec:FFT}
To further investigate the excitation of different azimuthal wavenumbers as a result of the nature of the inclined driver, we can apply a local Fourier decomposition to quantify the contribution of different azimuthal wavenumbers as a function of time at varying heights in our simulation. As the coronal loop in our setup does not have a defined boundary, denoted by some discontinuity in density or magnetic field, we instead choose to take azimuthal Fourier transforms at radial locations given by the flux surfaces of field lines $1$ and $2$. As the magnetic field expands with height to maintain total pressure balance, the radial position at which we conduct the azimuthal Fourier decomposition also varies with height. In other words, the radial location of the Fourier analysis on an individual field line is different in the corona than in the photosphere. Following the analysis conducted in \citet{ter2018} and \citet{magyar2022}, we adopt a discrete Fourier transform to analyze the contribution of the different azimuthal wavenumbers using the obtained profiles of $\hat{v}_r$ and $\hat{v}_{\varphi}$, for example those shown in Figure \ref{fig:vr_vphi_signals}. We use the notation $p$ rather than $m$ in order to distinguish between the excited azimuthal wavenumbers ($p$) in the simulation as opposed to the eigenmode solutions ($m$). It is then possible to write the discrete Fourier transform, namely, the function $g$, on each flux surface as:
\begin{equation}
    G(p) = \frac{1}{N} \sum_{k=0}^{N-1} g(k) e^{-i\frac{2\pi}{N}pk},
\end{equation}
for a discrete set of N samples ($p=0, ..., N-1$). In our case, the analysis is in the azimuthal direction, ranging from $0$ to $2\pi$, where the signals have been extended to cover a full azimuthal period as they only cover the range of $\varphi = [0, \pi]$ in the simulations. This means that instead of $k$ it is more convenient to introduce the parameter $\theta_k = 2\pi k/N$. The contribution of each excited wave mode $p$ to the total signal can be expressed, using the inverse Fourier transform, as:
\begin{equation}
    g(\theta_k) = \sum_{p=0}^{N-1} G(p) e^{ip\theta_k}.
\end{equation}
We note that, similar to \citet{ter2018} and \citet{magyar2022}, the spectrum of excited azimuthal modes is symmetric about $N/2$ and the Fourier transform of the signals $\hat{v}_r$ and $\hat{v}_{\varphi}$ are purely real and imaginary corresponding to cosine and sine components, respectively.

\begin{figure*}
\gridline{\fig{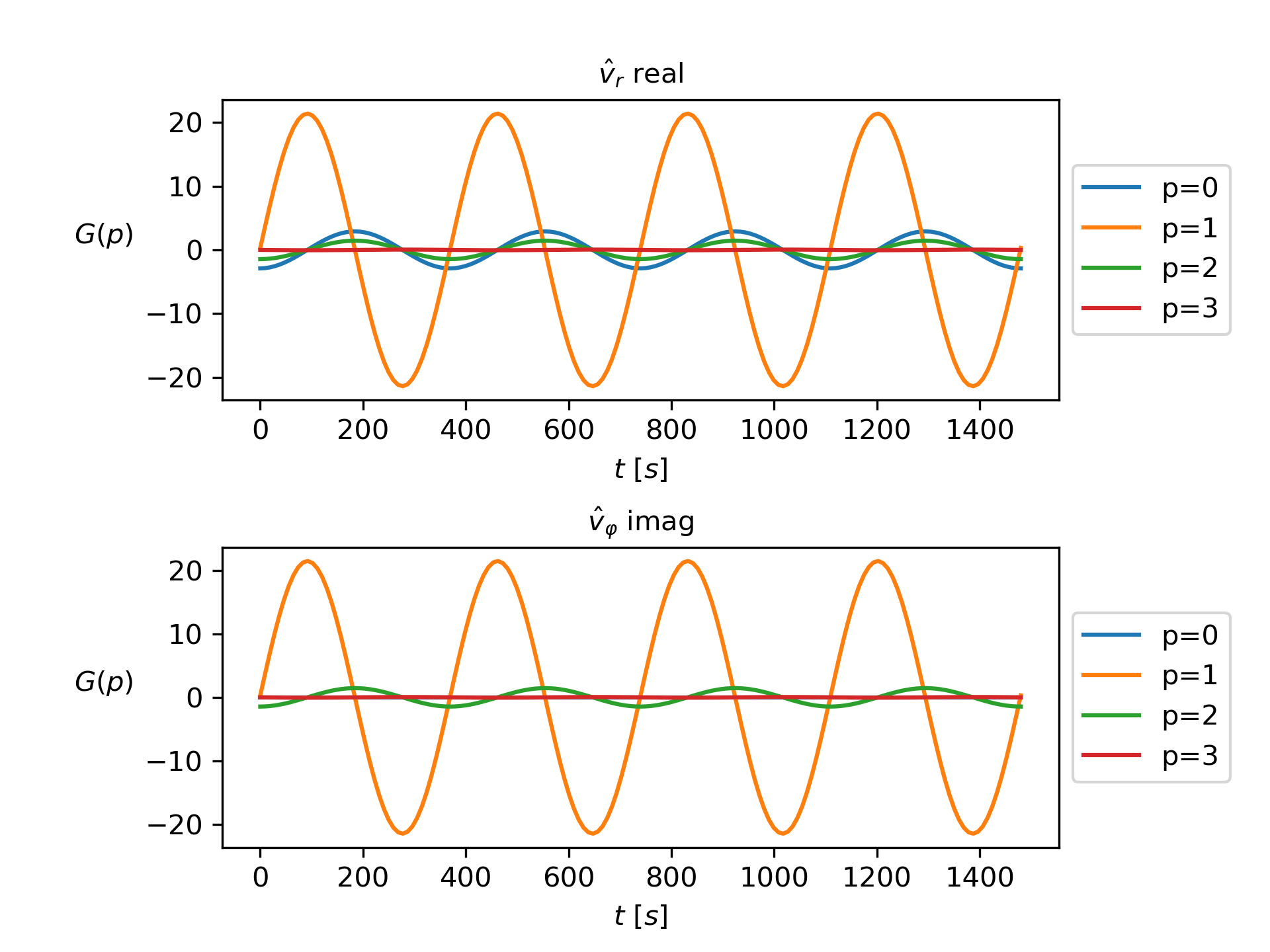}{0.49\textwidth}{(a) FFT at Field line 1 for driver}
          \fig{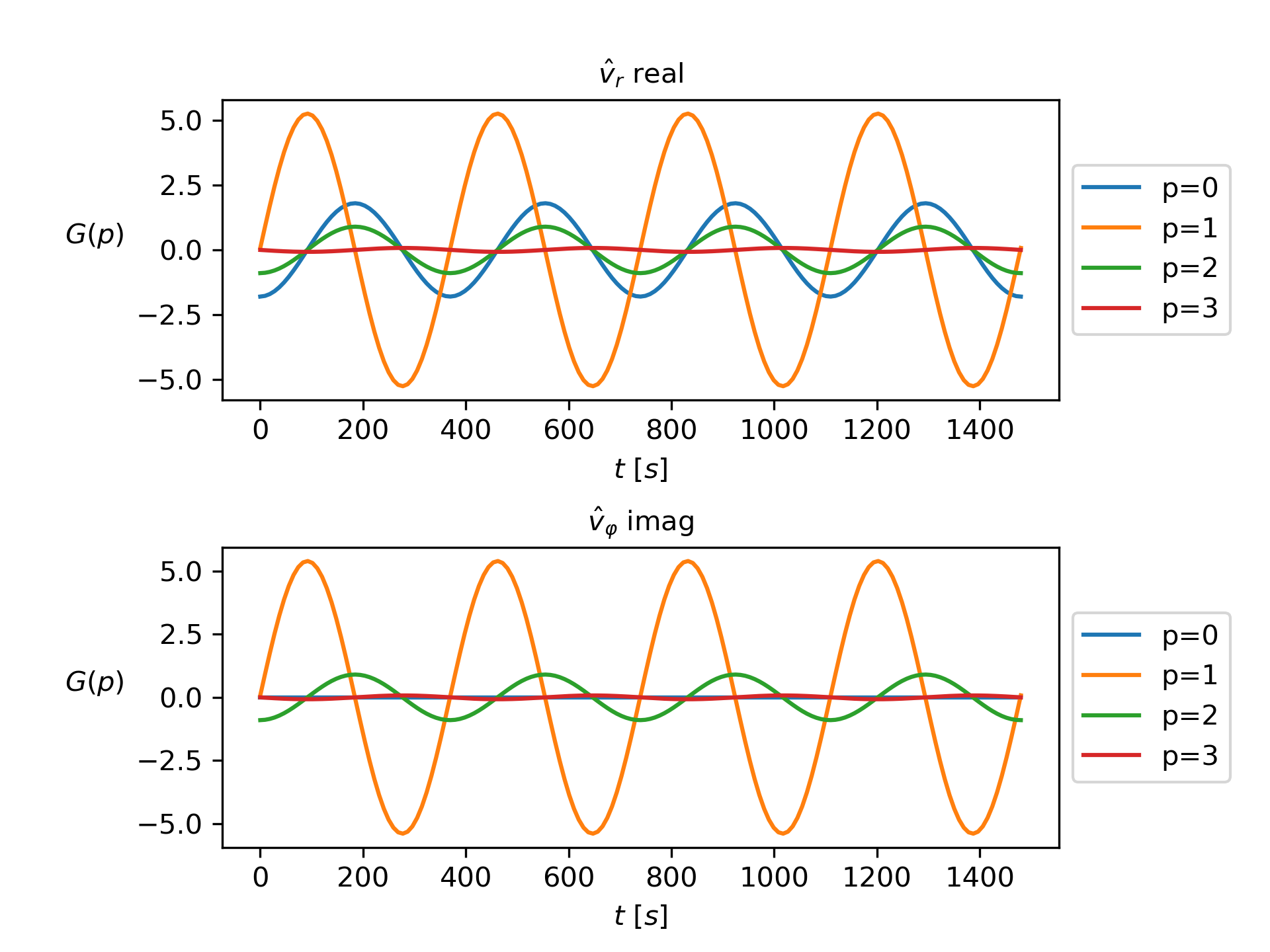}{0.49\textwidth}{(b) FFT at Field line 2 for driver}
          }
          
\gridline{\fig{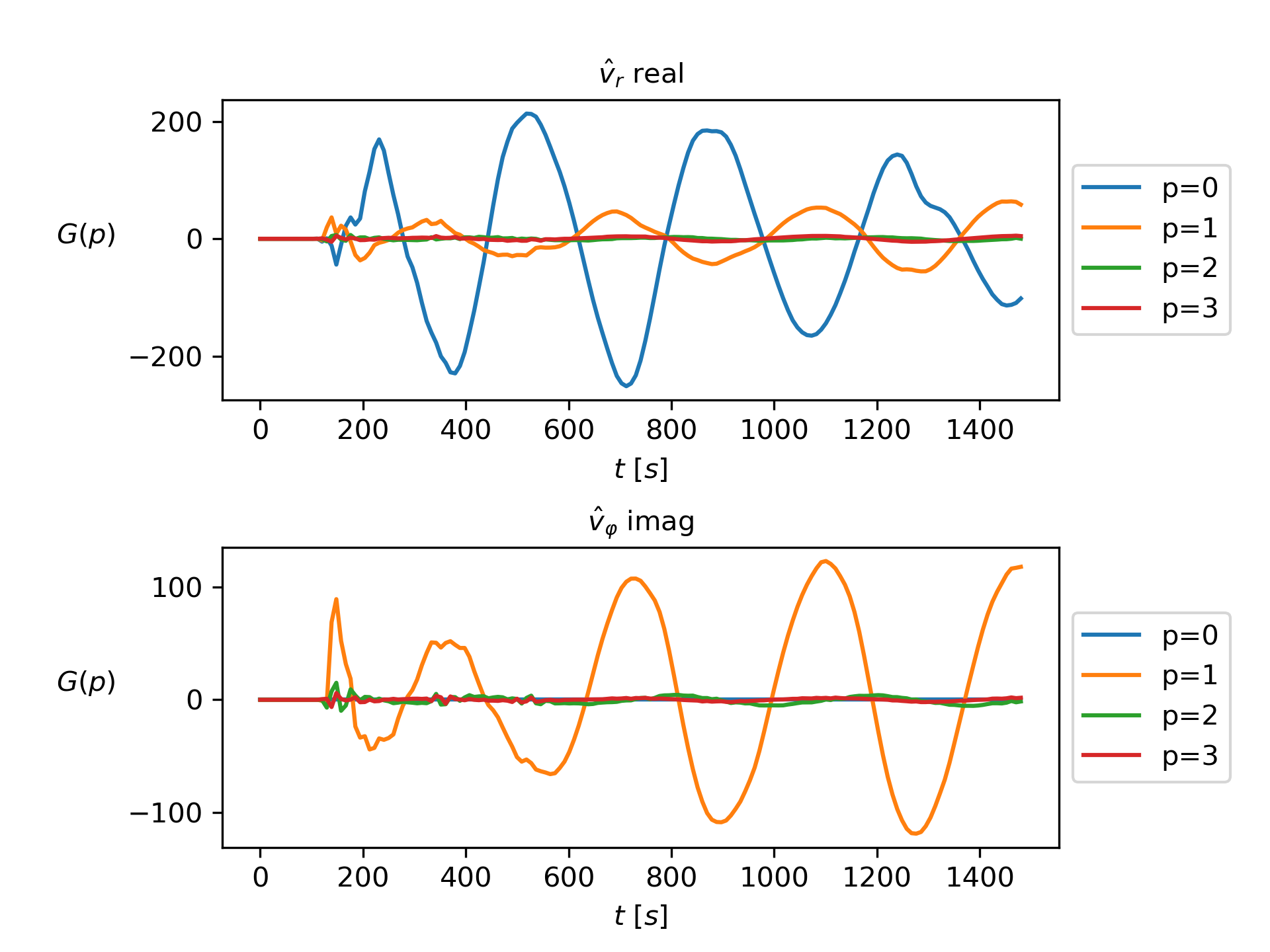}{0.49\textwidth}{(c) FFT on Field line 1 in photosphere}
          \fig{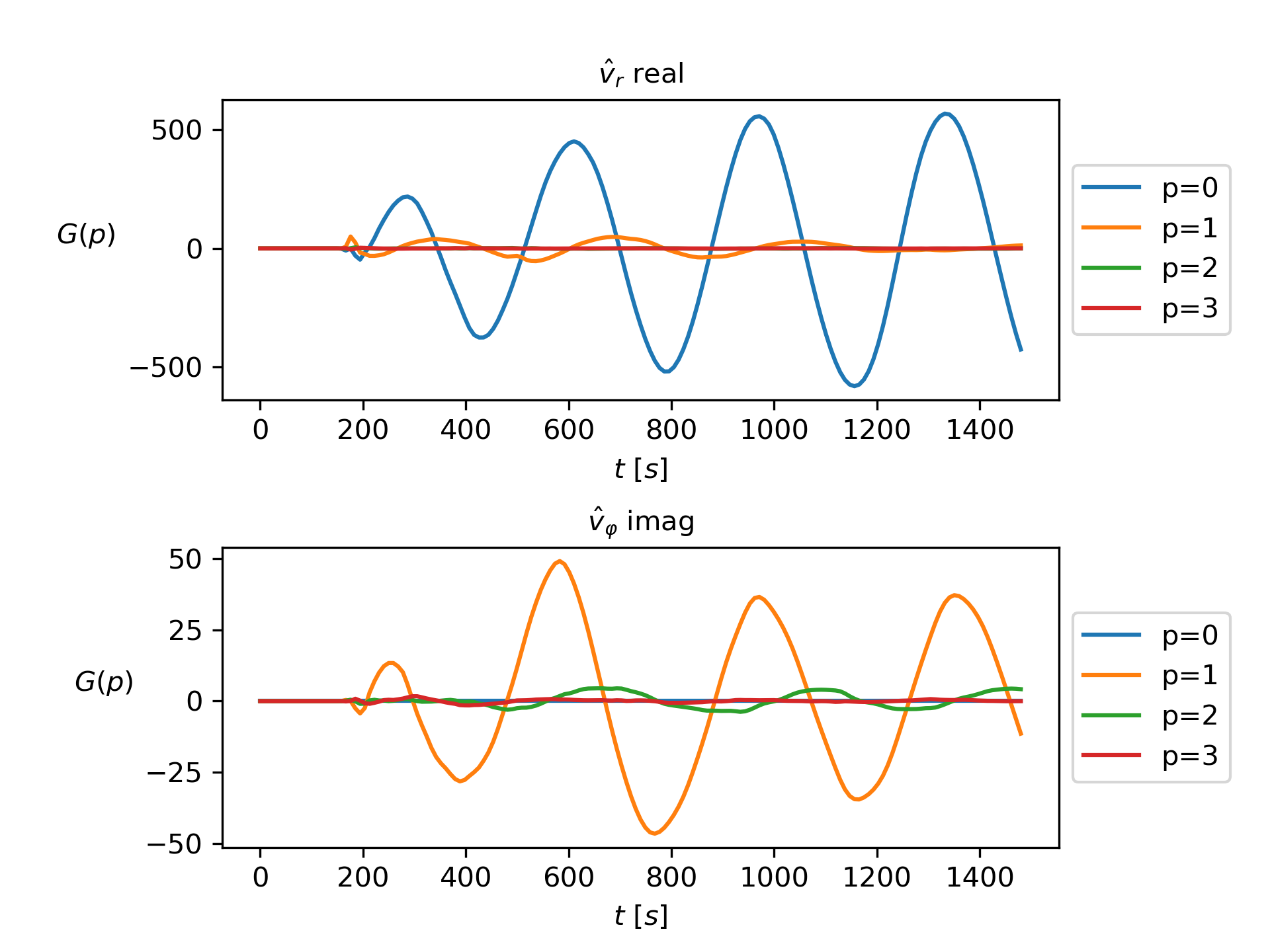}{0.49\textwidth}{(d) FFT on Field line 2 in photosphere}
          }

\gridline{\fig{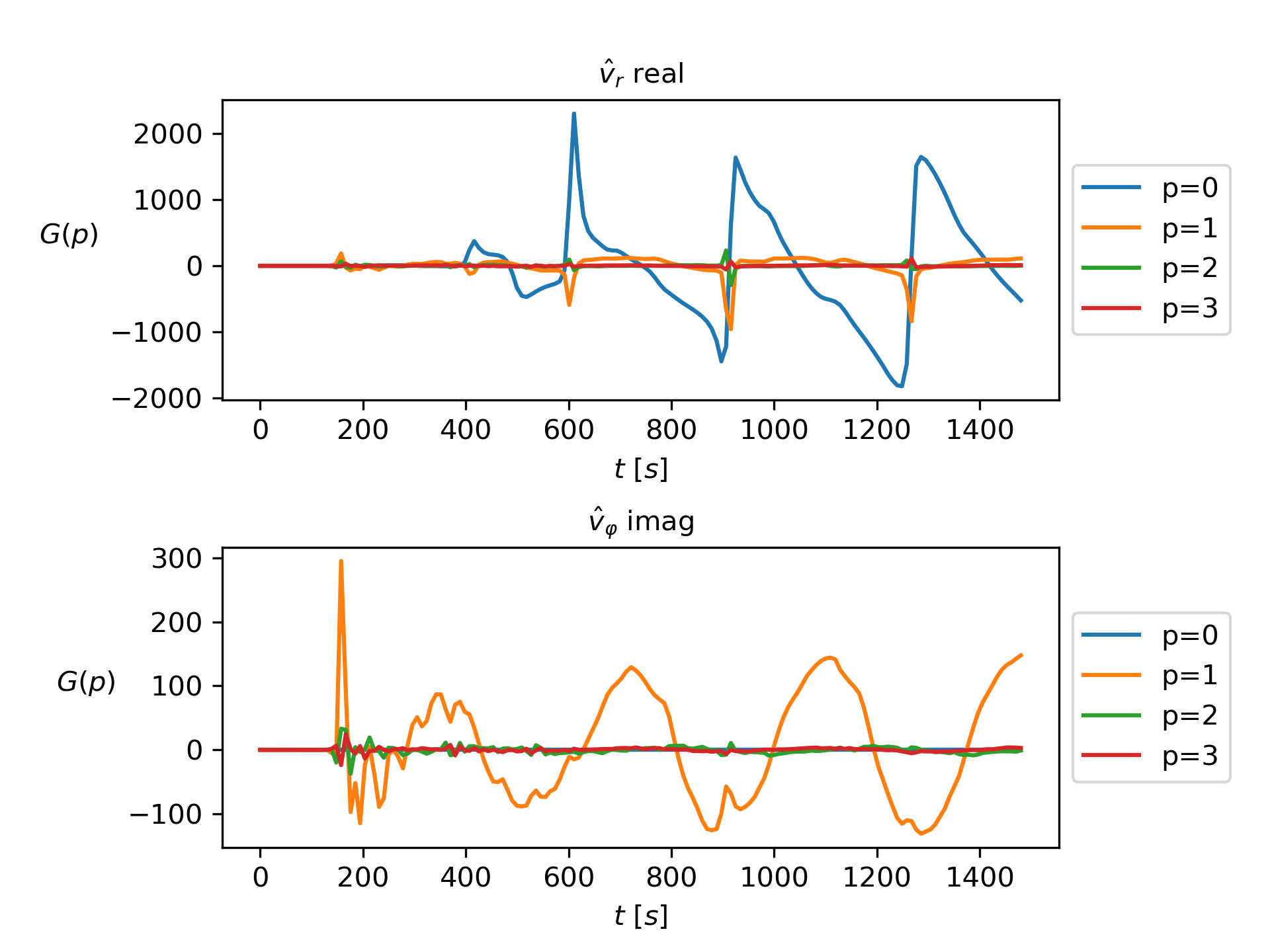}{0.49\textwidth}{(e) FFT on Field line 1 in corona}
          \fig{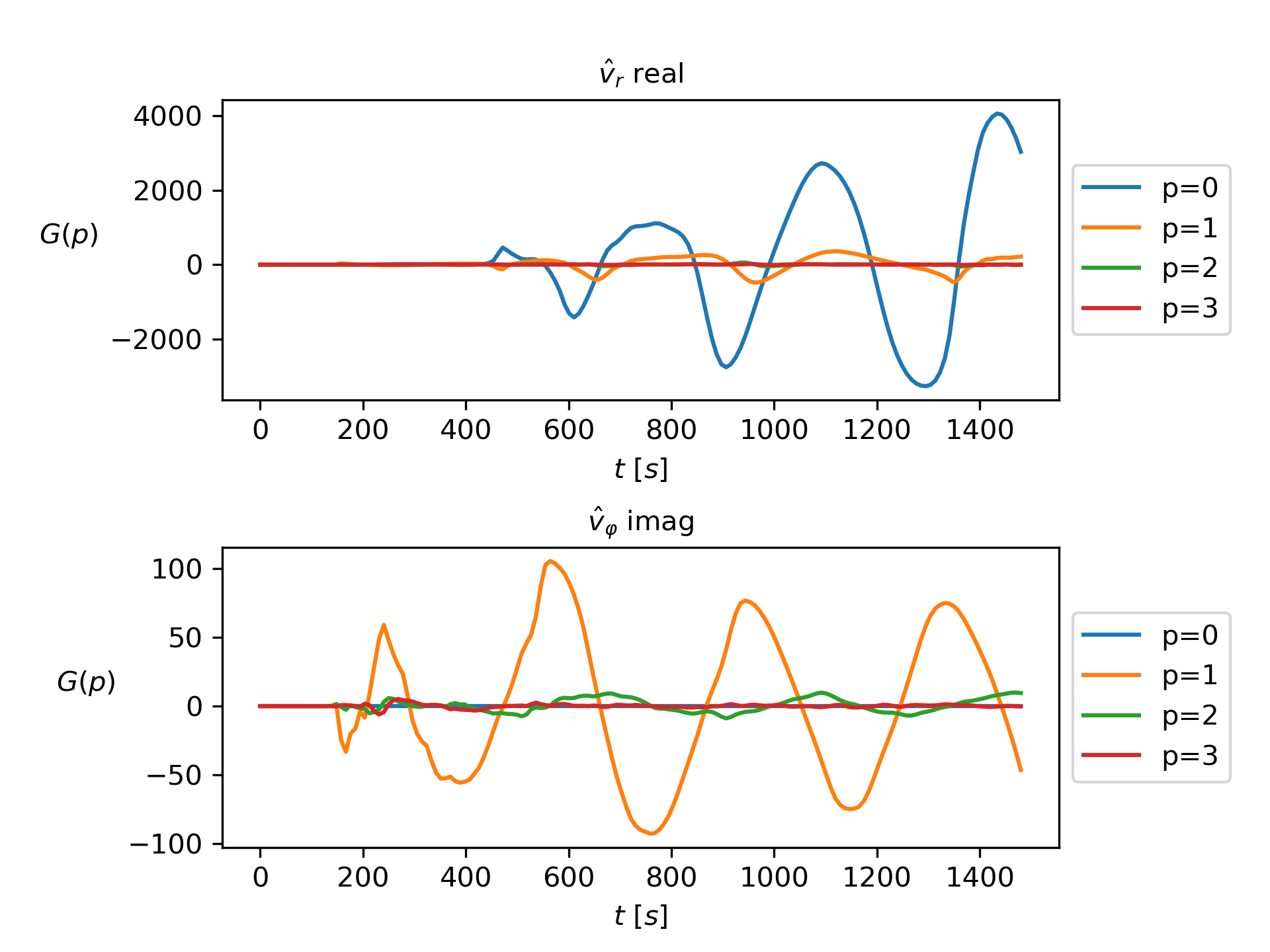}{0.49\textwidth}{(f) FFT on Field line 2 in corona}
          }
          
\caption{Fourier coefficients of $\hat{v}_r$ and $\hat{v}_{\varphi}$ in the photosphere ($z=-28.3$ Mm) and corona ($z=-23.0$ Mm) around flux surfaces at both field lines 1 (left) and 2 (right). Only the real component of the Fourier coefficient for $\hat{v}_r$ and the imaginary component for $\hat{v}_{\varphi}$ are shown corresponding to the cosine and sine contributions, respectively. The different azimuthal wavenumbers excited are represented by different colours, highlighted in the legend.
\label{fig:fourier_coefficients}}
\end{figure*}

The resulting Fourier analysis is displayed in Figure \ref{fig:fourier_coefficients} for the azimuthal Fourier coefficients on field lines 1 and 2 at both photospheric ($z=-28.3$ Mm) and coronal heights ($z=-23.0$ Mm). Furthermore, the Fourier coefficients associated with the driver are also computed at the footpoints of the respective field lines and shown in Figure \ref{fig:fourier_coefficients}. Firstly, we observe that, further away from the axis of the loop, there is increasing power in the sausage mode compared to the kink mode. This effect can also be seen in the contribution of the two modes at the location of the driver where, at larger radii, the amplitude of the driver is weaker as a result of the Gaussian profile with $\sigma = 2$ Mm (see Equation \ref{eq:Gaussian_driver}). This behaviour is expected as the amplitude of the inclined driver is reduced at larger radii. Therefore, the velocity amplitude of the driven waves decreases with increasing $r$, which results in a weaker contribution of magnetic tension (and magnetic pressure) to the restoring force of the waves. As a result of the amplitude of wave perturbation decreasing with radial distance from the loop, the driven transverse waves possess smaller velocity amplitudes further away from the locally applied driver.

The photospheric signals in Figure \ref{fig:fourier_coefficients} suggest that the driven waves convert into sausage modes in the magnetic loop, as is clearly seen in the Fourier coefficients of the $\hat{v}_r$ component, although there is also clearly some power in the kink modes, as anticipated from the previous section. It should also be noted here that there is no power in the $p=0$ axisymmetric modes from the analysis of the $\hat{v}_{\varphi}$ component, as expected because this mode does not produce azimuthal perturbations. Some power is present in higher order fluting modes ($p \geq 2$) however the power in these modes is negligible when compared to those corresponding to $p=0$ and $p=1$. 

Turning the attention now to the coronal signals, we see an interesting feature of the $\hat{v}_r$ Fourier coefficients. Along field line $1$, where the magnetic field is stronger compared to field lines at larger radii, the signals become very steep resembling a `saw-tooth' pattern, as a function of time, for all the modes shown in Figure \ref{fig:fourier_coefficients}. However, this feature is not seen at larger radii, for example in the signals displayed on field line $2$. A possible explanation for the steep $\hat{v}_r$ Fourier signals in Figure \ref{fig:fourier_coefficients}e could be an example of shock formation similar to that of umbral flashes \citep{Yuan2014, Houston2018, Anan2019}. Along field line $1$, the magnetic field is stronger and possesses a greater vertical component than along field line $2$. Although the field is less inclined closer to the centre of the waveguide, and also due to the nature of the driver, a $\hat{v}_r$ component is still present. However, as the field is less inclined, there is a stronger Alfv\'{e}n speed gradient along field line $1$. This gradient is reduced along field line $2$ due to the greater magnetic field inclination. As a result, the wave steepens along field line $1$ \citep[see e.g.][]{dep2004, Centeno2006, Yurchyshyn2014, Grant2018}. These signatures are not seen in the azimuthal signals as there is no stratification or magnetic field inclination/variation in this direction. Additionally, the acoustic cut-off region may play a role here as the inclined field lines reduce the cut-off frequency of waves propagating into the upper atmosphere \citep{Felipe2018,Felipe2020}, which may also be related to the cut-off frequency of specific tube waves \citep[e.g.][]{Spruit1981, Lopin2017, Pelouze2023}, although quantifying the contribution of the modified cut-off frequency is not within the scope of this study. Furthermore, the steep saw-tooth behaviour could also be an indication of a nonlinear interaction between the transition region and the upward propagating wave. The flux of the initial wave front is not transmitted into the corona, instead it is reflected and causes an oscillating wake of the transition region. The interaction with the moving transition region with the next wave front causes significant flux to be ejected into the corona at times $t=611$ s, $t=916$ s and $t=1267$ s, for example see Figure 5 of \citet{Riedl2021}, which corresponds to the times of the peak amplitudes in the saw-tooth signals along field line $1$. The dynamics of the transition region is not significantly affected by the wave fronts at larger radii, as the amplitude of the oscillating transition region has decreased, so this interaction between the transition region and the wavefront is not as strong and, as a result, the displayed Fourier coefficients are smoother in time.

\subsection{Vorticity}
For the case of a one-dimensional uniform straight cylinder with constant magnetic field \citep[e.g.][]{edrob1983}, parallel vorticity, namely the component of vorticity which is aligned with the magnetic field, is present as a delta function at the cylinder boundary. However, when the discontinuity in density is replaced with a continuous variation of density, \citet{goo2012} have demonstrated that vorticity is spread out over the whole interval with non-uniform density as a result of non-axisymmetric motions. Furthermore, recently, \citet{goo2019} have shown that in non-uniform plasmas MHD waves propagate both compression and parallel vorticity. Additionally, they have shown that the parallel, perpendicular, and radial components of displacement and vorticity are all non-zero.
\begin{figure*}
\gridline{\fig{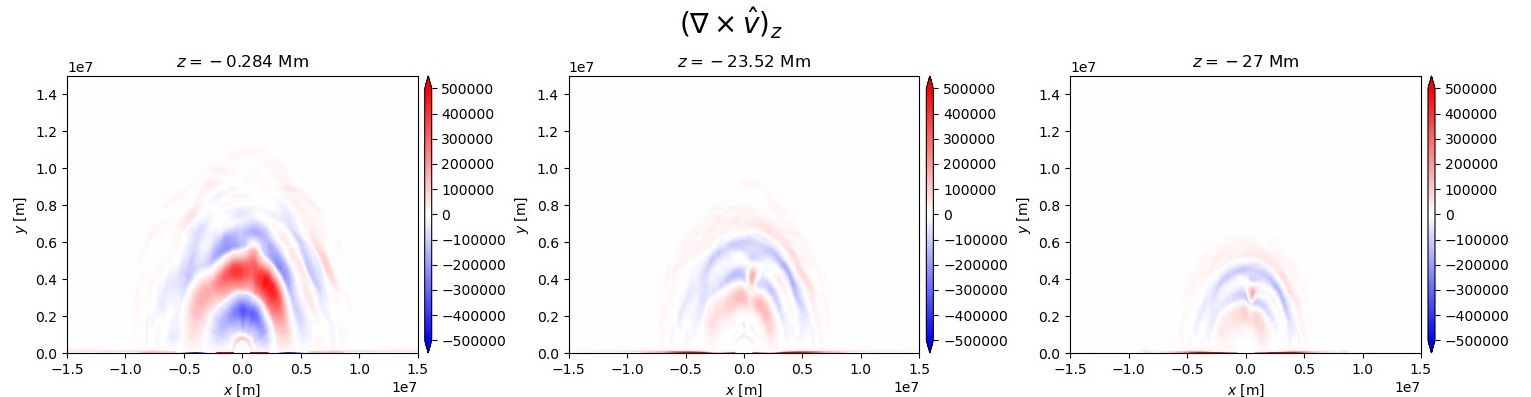}{0.99\textwidth}{}
          }          
\caption{The vertical component of the perturbed vorticity ($\nabla \times \hat{v}$)$_z$ at three different heights in the numerical domain including the loop apex (left), corona (middle) and chromosphere (right). 
\label{fig:vort_z_heights}}
\end{figure*}

Given the inhomogeneous nature of the atmosphere considered in our model, we would expect the non-axisymmetric motions to produce non-zero vorticity components that fill the whole non-uniform space. Figure \ref{fig:vort_z_heights} provides evidence of this by displaying the vertical component of vorticity at a given snapshot in time. The vertical vorticity is shown at three different heights corresponding to the apex of the structure (at $z \approx 0$ Mm), a coronal height slightly above the transition region and an additional height located below the transition region in the chromosphere. It is evident that the vertical vorticity occupies a greater portion of the domain at greater heights, where the magnetic field lines expand, compared to lower down in the atmosphere, where the plasma non-uniformity is provided through the magnetic field and density structuring. The vertical component of the vorticity is solely present as a result of the azimuthal motions, provided by the Alfv\'{e}nic nature of the excited waves and the breaking of azimuthal symmetry due to the inclined wave driver.

\begin{figure}
\gridline{\fig{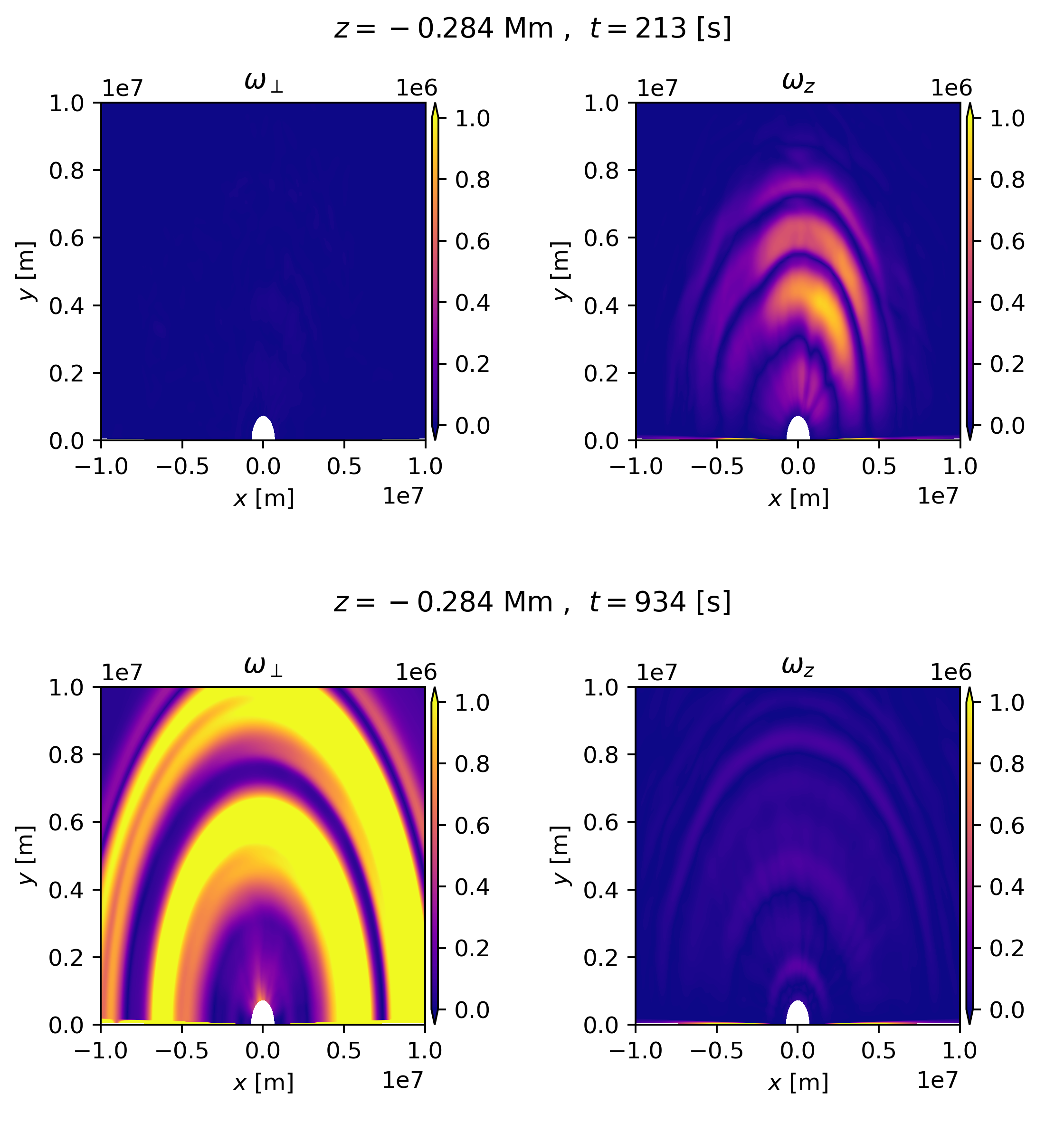}{0.485\textwidth}{}
          }        
\caption{The perpendicular and vertical component of the perturbed vorticity at the apex of the magnetic structure for two different times in the simulation. A time snapshot inside the initial wavefront ($t<370$ s) and at a later time ($t=934$ s). The colour-bar is scaled for all plots to make a comparison between the panels clearer.  
\label{fig:vort_comparison}}
\end{figure}

At the apex of the structure, the magnetic field lines are almost vertical, and the magnetic field inclination is negligible. As a result, the vertical vorticity can be approximated as the parallel vorticity and the vorticity perpendicular to the magnetic field can be approximated by $\omega_{\perp} = \sqrt{\omega_r^2 + \omega_{\varphi}^2}$. Figure \ref{fig:vort_comparison} displays the perpendicular and vertical vorticity at two times during the simulation. During the time period corresponding to the first wavefront from the driver, it is evident that the parallel vorticity is dominant over the perpendicular vorticity, which is the expected behaviour of a transverse wave and may be related to resonant absorption of a standing kink mode \citep{goo2011}. This is due to the coupling of transverse and azimuthal motions resulting in the amplitude of azimuthal motions increasing throughout the region of the non-uniform plasma \citep{Skirvin2022}. However, at a later time, the values of perpendicular vorticity significantly exceed those of the parallel vorticity, hinting that the longitudinal waves dominate the dynamics of the system. It is interesting to note that the initial wavefront of the driver produces transverse oscillations in the corona but there is no sign of parallel motions, for $t<370$ s, this can also be seen in the coronal signatures of Figure \ref{fig:fourier_coefficients}. The absense of signatures corresponding to parallel motions associated with slow modes in the corona during the first driven wave period agrees with the study of \citet{Riedl2021} and also the azimuthal wavenumber analysis presented in Section \ref{subsec:FFT}.

\section{Observational signatures of transverse motions}\label{sec:obs_features}

So far we have provided evidence that a driver at the base of a magnetic waveguide in a stratified solar atmosphere, such as a coronal loop, mimicking that of an inclined p-mode incident from below, can excite tube waves in the magnetic structure exhibiting similar properties to those associated with kink/Alfv\'{e}nic waves. It would be instructive to quantify these waves in terms of their observational characteristics, such that direct comparisons between observations and the numerical simulation presented here can be made.

\begin{figure}
\gridline{\fig{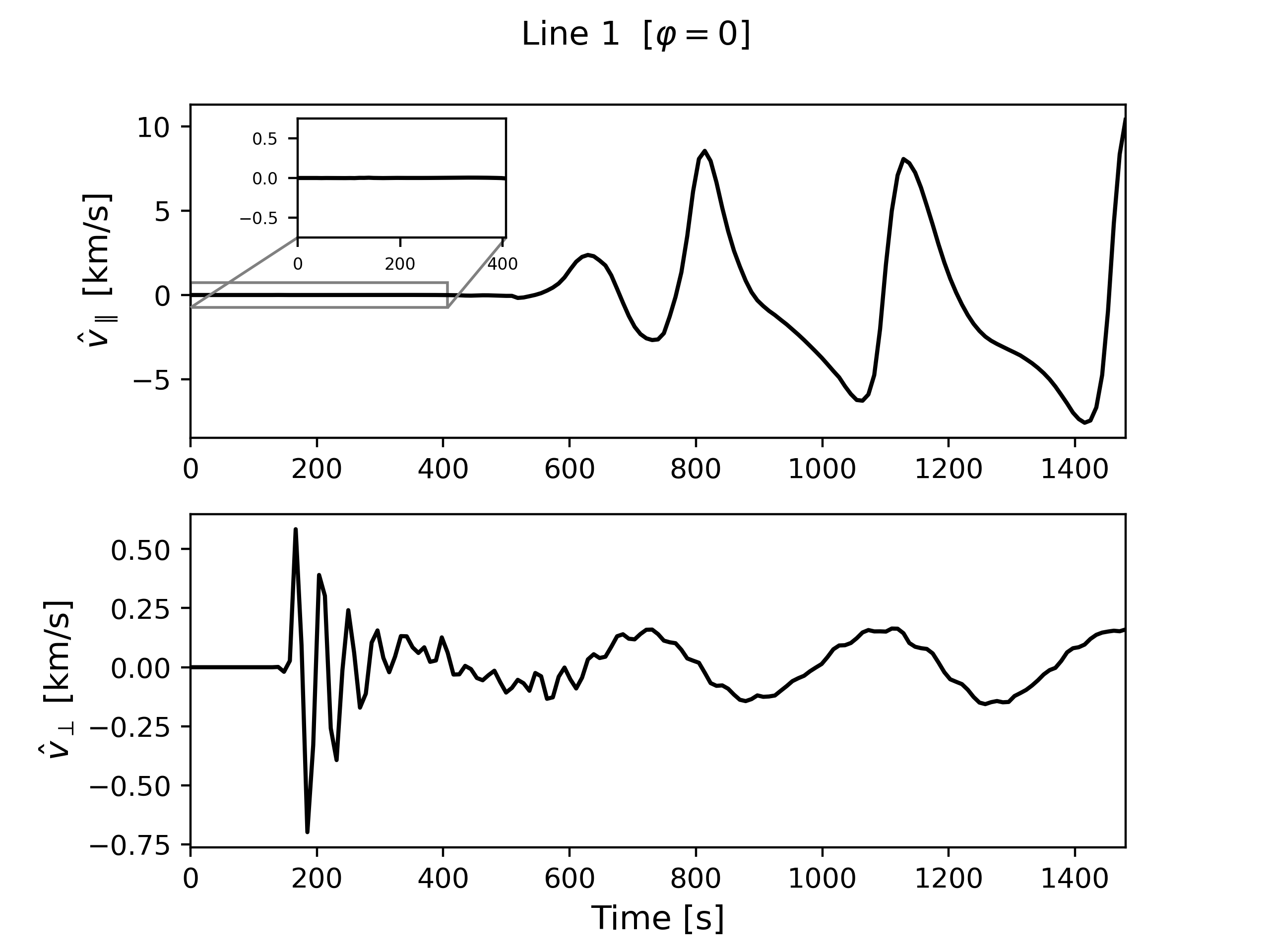}{0.485\textwidth}{}
          }        
\caption{The perturbed velocities parallel (top panel) and perpendicular (bottom panel) to the magnetic field at the apex of the loop on field line $1$ at an azimuthal slice corresponding to $\varphi=0$. A zoom-in plot is shown for the $\hat{v}_{\parallel}$ component such that a comparison with the perpendicular velocity can be made for earlier times in the simulation.
\label{fig:parvperp_vels}}
\end{figure}
Figure \ref{fig:parvperp_vels} highlights the differences between the parallel and perpendicular components of the perturbed velocity at the apex of the loop on field line $1$ at an azimuthal slice of $\varphi=0$. There is clearly an absence of parallel motions during the initial wavefront of the driven waves, whereas there are motions perpendicular to the magnetic field present, attributed to the presence of Alfv\'{e}nic waves. The magnitude of the transverse motions has a peak amplitude of $0.75$ km s$^{-1}$ which is smaller than some observational results in long coronal loops \citep{Anf2015,nak2016}, but lies within the range of the velocity amplitudes ($0.6-3.6$ km s$^{-1}$) of decayless oscillations detected in coronal bright points \citep{Gao2022}. Furthermore, the velocity amplitude in the simulation also roughly agrees with some results of previous spectroscopic observations \citep{Tian2012}. However, during the second, third and fourth driving periods, the amplitude of the transverse velocity reduces to around $0.2$ km s$^{-1}$, which is both smaller than the initial velocity and also the values reported in observations. However, this amplitude is sensitive to the strength of the driver and the strength of the magnetic field. The transverse oscillations also propagate in the corona with a phase speed comparable to the local Alfv\'{e}n speed which is compatible with Alfv\'{e}nic observations made using CoMP \citep{Tomczyk2009}. As expected, at later times, the motions parallel to the magnetic field dominate due to the presence of slow modes propagating at the local sound speed. However, the amplitude of these motions in the corona could potentially be damped by non-ideal processes, such as thermal conduction, which is expected to be important at greater heights where the temperature increases significantly \citep{VanDoorsselaere2011, KrishnaPrasad2018, Duck2021}, although this aspect is not considered in our current model. 

\begin{figure*}
\gridline{\fig{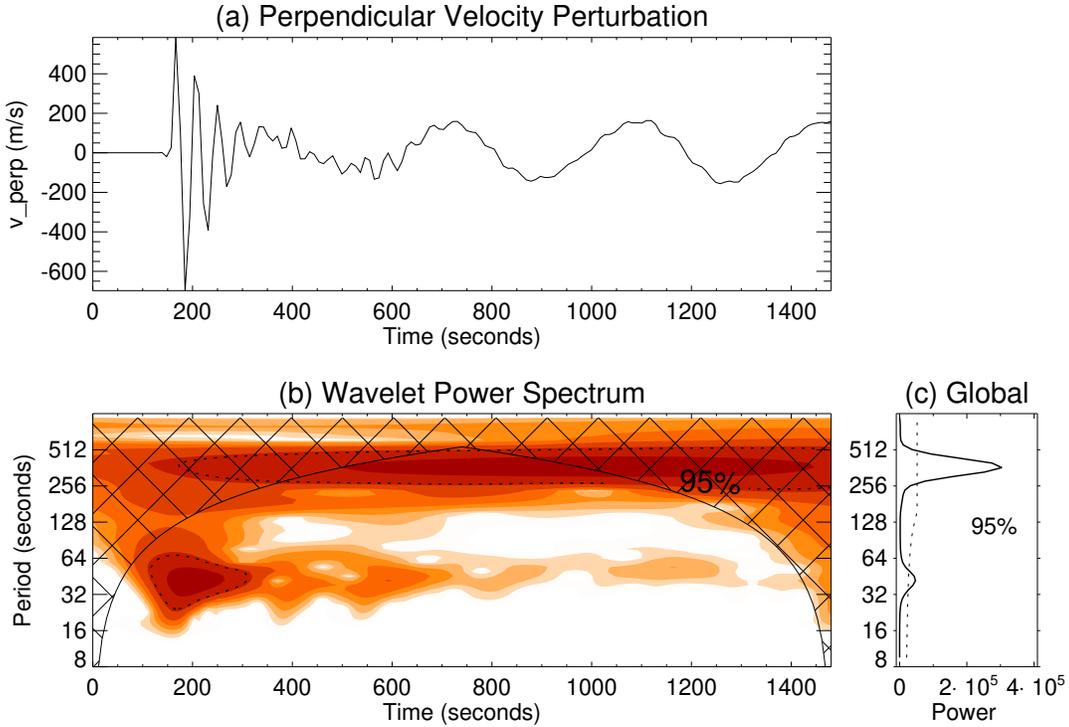}{0.8\textwidth}{}
          }        
\caption{(a): Time series of the perpendicular velocity perturbation at loop apex ($z=0$) on field line 1. (b): Wavelet spectrum of the time series, with darker colours representing stronger powers. (c): Global wavelet spectrum of the time series. The dashed lines in (b) and (c) represent a significance level of 95\%.
\label{fig:wavelet}}
\end{figure*}

Additionally, we can investigate the periods of the transverse oscillations present in the simulation. We repeat the time series of the velocity perturbation perpendicular to the loop axis at loop apex ($z=0$) on field line 1, shown in Figure \ref{fig:wavelet} (a).
The case for field line 2 is similar however the amplitude of the perturbed motions is decreased. It can be seen that there is initially a short-period oscillation, followed by a long-period oscillation. To further investigate the property of these two oscillation regimes, we can conduct a wavelet analysis, which is presented in Figure \ref{fig:wavelet} (b) and (c). The wavelet power peaks at two regions in the spectrum, corresponding to the short-period oscillation in the beginning and the long-period oscillation throughout the whole simulation time, respectively. We can obtain the peak periods from the global wavelet spectrum, which are $42$ s and $364$ s. The latter value is close to the period of our p-mode driver ($370$ s), which suggests that the long-period oscillation is directly driven by p-modes. As for the oscillation with a period of $42$ s, we believe that this may correspond to the eigenmode of the loop. The presence of these two distinct oscillation regimes is interesting, however a detailed analysis of this is not the aim of the present study and will be addressed in future work.

\section{Conclusions}\label{sec:conclusions}

In this study, we have shown that an acoustic-gravity wave driver implemented at the photosphere, mimicking that of solar p-modes, can produce Alfv\'{e}nic perturbations in the corona when waves are driven at an angle to the vertical axis of a magnetic structure, such as a coronal loop. The inclination of the driver in this study was taken to be $\theta = 15^{\circ}$, which generates predominantly vertical waves, however, it breaks the azimuthal symmetry of the system. By extending previous studies, we have performed a numerical simulation using the PLUTO code and analysed the perturbations of the resulting modes in the framework of a cylindrical flux tube model. We have demonstrated that the axis of the waveguide is perturbed as a result of the wave propagation, which can be explained through the presence of the $m=1$ kink mode, however, due to the non-uniformity of the plasma in our model it may be more appropriate to refer to these resulting motions as `Alfv\'{e}nic'. Furthermore, by performing a Fourier analysis in the azimuthal direction of the domain, we have quantified the contribution of different modes to the perturbed velocity signals. We have shown that the cylindrical sausage mode is the dominant mode in the flux tube (due to the nature of the driver), however, a clear contribution of the kink mode to the perturbed signals can be seen, in addition to small contributions from higher order fluting modes. The strength of the kink mode decreases with distance from the central driver as the restoring force of magnetic tension becomes weaker. We have shown that the coronal signals display a behaviour which is commonly associated with shock formation, by the development of a saw-tooth pattern in their velocity signals as a function of time. This behaviour is seen in the radial component of the perturbed velocity, however, not in the azimuthal component. This is associated to the inclination of the magnetic field lines in the gravitationally stratified model. We also computed the perturbed vorticity to demonstrate that the Alfv\'{e}nic fluctuations produce field-aligned vorticity, which occupies the whole non-uniform space, as predicted from theory. Additionally, the vorticity analysis identified two separate vorticity regimes, where the ratio of the parallel to the perpendicular vorticities change, due to the difference in propagation speeds between the longitudinal and transverse waves. Finally, we have shown that the Alfv\'{e}nic waves with velocity amplitudes in the range of $0.2 - 0.75$ km s$^{-1}$ display two different regimes of wave period corresponding to $42$ s and $364$ s. We attribute these periods to the eigenmode of the magnetic loop and the waves related directly to the driver, respectively. The presence of oscillations with concurrent wave periods will be the focus of a future study. 

We have presented a model of an inclined acoustic driver incident on a vertical magnetic structure, which, for small values of inclination, is similar to a model of a purely vertical driver incident on an inclined loop. In the specific scenario where the model would not include gravitational stratification, this setup would be identical to a vertical driver incident on an inclined magnetic loop. Although, as a result of gravitational stratification, field-guided MHD waves, such as Alfv\'{e}n waves and magnetoacoustic waves in specific plasma-$\beta$ regimes, travel further distances through the vertically stratified lower atmosphere as opposed to those waves travelling along a purely vertical field and experience a modified effective gravity along the field. Therefore, care should be taken when modeling inclination angles that are sufficiently large (i.e. $>30^{\circ}$), as the correction to the field-aligned gravity may produce a non-negligible influence on the results. It would be an interesting future study, with relevance to observations of the solar atmosphere, to model such acoustic waves incident on an inclined magnetic structure in a stratified solar atmosphere. Additional avenues of future studies include a detailed analysis of mode conversion occurring in the lower atmosphere of the simulation, as the initial wave driver is predominantly acoustic in nature, however, the driven waves clearly convert to tube waves and develop magnetic properties, which is expected to occur around the equipartition layer \citep{Khomenko2012}. Furthermore, there is evidence of phase mixing developing in the simulation, due to the cross-field structuring, which may display observable signatures \citep{Kaneko2015}. A more detailed discussion regarding the energetics of the excited waves should be undertaken in order to quantify their role in the energy budget of the solar atmosphere and how they may contribute to the coupling of various atmospheric layers.

\acknowledgments

We are grateful to the anonymous referee for their constructive and fruitful suggestions. SJS and TVD were supported by the European Research Council (ERC) under the European Union's Horizon 2020 research and innovation programme (grant agreement No 724326) and the C1 grant TRACEspace of Internal Funds KU Leuven. The results received support from the FWO senior research project with number G088021N. YG was supported by China Scholarship Council under file No. 202206010018.

\bibliography{ref}{}
\bibliographystyle{aasjournal}

\end{document}